\documentclass[sigconf,authorversion,nonacm]{acmart}

\begin{document}

%%
%% The "title" command has an optional parameter,
%% allowing the author to define a "short title" to be used in page headers.
\title{COVID19-CBABM: A City-Based Agent Based Disease Spread Modeling Framework}

%%
%% The "author" command and its associated commands are used to define
%% the authors and their affiliations.
%% Of note is the shared affiliation of the first two authors, and the
%% "authornote" and "authornotemark" commands
%% used to denote shared contribution to the research.
\author{Raunak Sarbajna}
\authornote{All authors contributed equally to this research.}
\email{rsarbajna@uh.edu}
% \orcid{1234-5678-9012}
\author{Karima Elgarroussi}
\authornotemark[1]
\email{karima.elgarroussi@gmail.com}
\author{Hoang D Vo}
\authornotemark[1]
\email{hoangvod@gmail.com}
\affiliation{%
 \institution{University of Houston}
 \city{Houston}
 \state{Texas}
 \country{USA}}

% \author{Lars Th{\o}rv{\"a}ld}
% \affiliation{%
%   \institution{The Th{\o}rv{\"a}ld Group}
%   \streetaddress{1 Th{\o}rv{\"a}ld Circle}
%   \city{Hekla}
%   \country{Iceland}}
% \email{larst@affiliation.org}

% \author{Valerie B\'eranger}
% \affiliation{%
%   \institution{Inria Paris-Rocquencourt}
%   \city{Rocquencourt}
%   \country{France}
% }

% \author{Karima Elgarroussi}
% \affiliation{%
%  \institution{University of Houston}
% %  \streetaddress{Rono-Hills}
%  \city{Houston}
%  \state{Texas}
%  \country{USA}}
% \email{karima.elgarroussi@gmail.com}

\author{Jianyuan Ni}
\affiliation{%
 \institution{Juniata College}
%  \streetaddress{https://orcid.org/0000-0002-6968-6536}
 \city{Huntingdon}
 \state{Pennsylvania}
 \country{USA}}
\email{jni100@juniata.edu}

\author{Christoph F. Eick}
\affiliation{%
 \institution{University of Houston}
%  \streetaddress{Rono-Hills}
 \city{Houston}
 \state{Texas}
 \country{USA}}
\email{CEick@.uh.edu}

% \author{Hoang D Vo}
% \affiliation{%
%  \institution{University of Houston}
% %  \streetaddress{Rono-Hills}
%  \city{Houston}
%  \state{Texas}
%  \country{USA}}
% \email{hoangvod@gmail.com}
% % \email{rsarbajna@uh.edu}

% \author{Huifen Chan}
% \affiliation{%
%   \institution{Tsinghua University}
%   \streetaddress{30 Shuangqing Rd}
%   \city{Haidian Qu}
%   \state{Beijing Shi}
%   \country{China}}

% \author{Charles Palmer}
% \affiliation{%
%   \institution{Palmer Research Laboratories}
%   \streetaddress{8600 Datapoint Drive}
%   \city{San Antonio}
%   \state{Texas}
%   \country{USA}
%   \postcode{78229}}
% \email{cpalmer@prl.com}

% \author{John Smith}
% \affiliation{%
%   \institution{The Th{\o}rv{\"a}ld Group}
%   \streetaddress{1 Th{\o}rv{\"a}ld Circle}
%   \city{Hekla}
%   \country{Iceland}}
% \email{jsmith@affiliation.org}

% \author{Julius P. Kumquat}
% \affiliation{%
%   \institution{The Kumquat Consortium}
%   \city{New York}
%   \country{USA}}
% \email{jpkumquat@consortium.net}

%%
%% By default, the full list of authors will be used in the page
%% headers. Often, this list is too long, and will overlap
%% other information printed in the page headers. This command allows
%% the author to define a more concise list
%% of authors' names for this purpose.
% \renewcommand{\shortauthors}{Trovato and Tobin, et al.}

%%
%% The abstract is a short summary of the work to be presented in the
%% article.
\begin{abstract}
 In response to the ongoing pandemic and health emergency of COVID-19, several models have been used to understand the dynamics of virus spread. Some employ mathematical models like the compartmental SEIHRD approach (susceptible-exposed-infectious-hospitalized-recovered-dead) and others rely on agent-based modeling (ABM). In this paper, a new city-based agent-based modeling approach called COVID19-CBABM is introduced. It considers not only the transmission mechanism simulated by the SEHIRD compartments but also models people's movements and their interactions with their surroundings, particularly their interactions at different types of Points of Interest (POI), such as supermarkets. Through the development of knowledge extraction procedure for Safegraph data, our approach simulates realistic conditions based on spatial patterns and infection conditions considering locations where people spend their time in a given city. Our model was implemented in Python programming language using the Mesa-Geo framework. COVID19-CBABM is portable and can be easily extended by adding more complicated rules/scenarios. Therefore, it is a useful tool to assist the government and health authorities in evaluating strategic decisions and actions efficiently against this epidemic, using the unique mobility patterns of each city.
\end{abstract}

%%
%% The code below is generated by the tool at http://dl.acm.org/ccs.cfm.
%% Please copy and paste the code instead of the example below.



%%
%% Keywords. The author(s) should pick words that accurately describe
%% the work being presented. Separate the keywords with commas.

%% A "teaser" image appears between the author and affiliation
%% information and the body of the document, and typically spans the
%% page.
% \begin{teaserfigure}
%   \includegraphics[width=\textwidth]{sampleteaser}
%   \caption{Seattle Mariners at Spring Training, 2010.}
%   \Description{Enjoying the baseball game from the third-base
%   seats. Ichiro Suzuki preparing to bat.}
%   \label{fig:teaser}
% \end{teaserfigure}

%%
%% This command processes the author and affiliation and title
%% information and builds the first part of the formatted document.
\maketitle
\section{Introduction} \label{sec:intro}

The new ongoing pandemic of COVID-19 has spread all over the world threatening the lives of many people. The World Health Organization (WHO) declared the outbreak a Public Health Emergency of International Concern on January 30th, 2020, and a pandemic on March 11th. The first laboratory-confirmed case of COVID-19 in the United States was confirmed on January 20, 2020 and reported to CDC on January 22, 2020. In late February, there were just a dozen known cases in the US—most of them linked to travel. But by summer, the virus started to infect more and more people: As of August 19th, 2020, the country reached more than 5,509,776 confirmed cases and more than 172,204 deaths by COVID-19, according to the CDC which makes the US the first epicenter of the virus in the world. Thus, in March 2020, several health precautions and governmental strategies were taken to control the COVID-19 outbreak including closing schools and points of interest and implementing the social distancing plan. These measures resulted in "lowering the curve" in many states; however, the virus continued to affect every part of the American cities with an enormous growth of cases.

 Given the unavailability of an actual cure/vaccine for this virus, simulating and modeling the pandemic is a crucial tool to understand its dynamics, measure its epidemiological effects and most importantly control its spread. Following this line of thought, many researchers have developed mathematical models like the SIR(Susceptible-Infected-Recovered) [22][23] or SEIR (Susceptible-Exposed-Infected-Recovered) models [24][25] to predict the spread of COVID-19. However, these models are non-spatial approaches as they only consider the transmission mechanism among the population as a whole and ignore spatial variations in human interactions as well as diversity in the human population. To overcome these limitations, agent-based simulation models (ABM) [7][20] have also been proposed: Due to their simplicity of implementation and flexibility, ABM have the capability to simulate the dynamics of complex systems by capturing the heterogeneity of its population and predicting the global effects that emerge due to the interactions between the agents over time and space.
      
We note that recent studies using mathematical models or agent-based models solely model the spread of the infectious disease either through differential equations of the infection rates or through hypothetical simulations \cite{cuevas_agent-based_2020}. Specific social distancing practices by citizens are ignored as well in the standard SEIR model. In this work, we propose a city-based, agent-base model \cite{van_dyke_parunak_agent-based_1998} called COVID19-CBABM to simulate and predict the spread of COVID-19 by using the Bronx borough of New York City (NYC) as a test case. COVID19-CBABM uses a geospatial approach to emulate people’s mobility and interactions with their surroundings paying special attention to points of interest (POIs) in a city, such as restaurants, supermarkets, and parks. It takes advantage of both SEIHRD and agent-based models’ capabilities by using different compartments, where the population of agents can interact following specific rules. We are simulating two types of agents including the individuals and the POIs. The individuals are assumed to move to selected POIs following an assigned schedule. 

To the best of our knowledge, there has been no widely accepted model of disease spread within a specific city using actual mobility patterns with respect to POIs of its citizens. Our model, COVID19-CBABM, allows to predict infection and death rates for a particular city. We feel that this model will help the government efficiently prescribe local interventions as the severity of COVID-19 changes, according to each city’s governmental and health measures. SafeGraph data is used to obtain realistic POI parameters. The parameter extraction methods we develop can be reused when developing similar models for other cities which makes our  COVID19-CBABM portable and extendable.

The main contributions of this paper include:
\begin{itemize}
    \item A modular, city-based, agent based epidemiological model that can deal with spatial variation.
    \item Automated extraction of model parameters from existing public datasets.
    \item Design of a portable and reusable epidemiological model.
    \item An epidemiological model that focuses on modelling the interaction of citizens with Points-of-Interests.
\end{itemize}

The remainder of this paper is organized as follows: Section II presents in detail the architecture of the COVID19-CBABM. Section III describes the experiment setup, the datasets that we use, our evaluation method, and the results of the experiment that has been performed to demonstrate the effectiveness of our model. Finally, section IV concludes the paper and suggests directions for future work.

\section{Related Work } \label{sec:rel_work}

After the Novel Coronavirus Disease, COVID-19 was declared as a pandemic by the World Health Organization on March 11, 2020, researchers and scientists around the globe investigated the dynamics of the SARS-CoV-2-type viruses using different approaches.  The equation and the agent-based models are the most popular in the literature.  In the equation-based modeling (EBM), the model is a set of equations, and the execution consists of evaluating them [11]. The SEIR (Susceptible-Exposed-Infectious-Recovered) model developed by Kermack and McKendrick [1] is one of the most adopted mathematical models in this category. Recently, it has been proposed in different contexts to model and analyzes the spread of COVID-19 [3][2][6][8][9][10]. In other research, new states have been added to improve the simulation of the model such as the dead compartment in [12][17]. Ndairou et al. considered an extra class of super-spreaders in every compartment [13], while other work takes the undetected infections [14] asymptomatic infected groups [15], and hospitalization [16] into consideration.
      
On the other hand, the agent-based models (ABM) are becoming popular in infectious disease epidemiology as they can capture complex dynamics of disease spread that other types of models cannot [18]. Generally, they consist of a group of agents interacting with each other in a shared environment. The behavior of those agents is governed by a set of rules within a loop where the agents run and interact.  These microsimulation models have been repurposed in some papers to simulate the spread of COVID-19 transmission. Cuevas et al. [20] propose an agent-based model to evaluate the COVID-19 transmission risks in facilities focusing on the importance of individual contact patterns in the modeling. Bossert et al. [7] developed an agent-based model combining socio-economic and traffic data to analyze COVID-19 spreading in a South Africa city under social isolation scenarios. Hoertel et al. [19] developed a stochastic agent-based microsimulation model of the COVID-19 epidemic in France, where they examined the potential impact of post-lockdown measures on cumulative disease incidence and mortality, and on intensive care unit (ICU)-bed occupancy. De Falco et al. [21] extend the SEIR model by including modern social distancing practices as used in two Italian regions and analyzed their impact on various scenarios.

In 2021, Kerr et al. [16] developed the ABM named Covasim to produce a robust system that could apply accurate policy decisions in simulating COVID-19 spread in countries. Covasim is an open-source model that includes demographic information on age structure, population size, social distance, schools, workplaces, and hygiene measures [16]. Covasim utilizes advanced software tools and computational methods to minimize the complexity and computation time of running ABMs. Covasim has been used to inform policy decisions in the United States, Vietnam, the United Kingdom, Australia, India, Russia, Kenya, and South Africa [16].

To the best of our knowledge, no work in the literature uses agent-based models to simulate the epidemiological impacts of the COVID-19 relying on realistic mobility data in a specific city. In this work, we are interested in analyzing how the dynamics of COVID-19 infection can emerge from individuals’ activities with respect to POIs they visit. Hence, this paper proposes a simple COVID19-CBABM in which we reused the 6 compartments of the SEIHRD model to characterize the states of our agents and analyzed different scenarios of social distancing for the Bronx borough while resolving the scalability and parameter selection challenges associated with the agent-based model.

\section{SEIR-CABM Model Overview} \label{sec:mod_ov}

\section{SEIR Models}

COVID19-CBABM utilizes an Agent-Based Model framework in which individual agents follow the compartmental epidemiological model called SEIHRD to understand and predict the spread of the COVID-19 disease caused by the novel coronavirus within Bronx borough. Relying on the SEIHRD framework, the COVID19-CBABM aims to forecast the infection rate and death rate of COVID-19 in a city. 

The basic SEIHRD model has six components, which represent at any given point of time the number of susceptible individuals (S), the number of exposed individuals (E), the number of infectious individuals (I), the number of hospitalized individuals (H), the number of recovered individuals (R), and the number of dead individuals (D). A SEIHRD model relies on the following non-negative, real-valued parameters:

\begin{itemize}
    \item $\alpha$: The inverse of the incubation period of the disease.
    \item $\beta$: The average contact rate of the population in question.
    \item $\gamma$: The inverse of the mean infectious period.
    \item $\delta$: The death rate of the population caused by COVID-19
\end{itemize}

Generally, SEIHRD models are solved through Differential Evolution. This involves creating an initial set of random possible solutions to the problem, with the size of the initial population kept constant during the evolution. To make the simulation realistic, the model makes some assumptions to decide the movements of individuals agents as follows: 1) the total number of Individual agents remains unchanged during the simulation; 2) Individual agents cannot travel further than a certain distance within one iteration; 3) an R agent never becomes an I agent.

\section{Model Architecture}
\begin{figure}[t]
\centering
\includegraphics[width=\columnwidth]{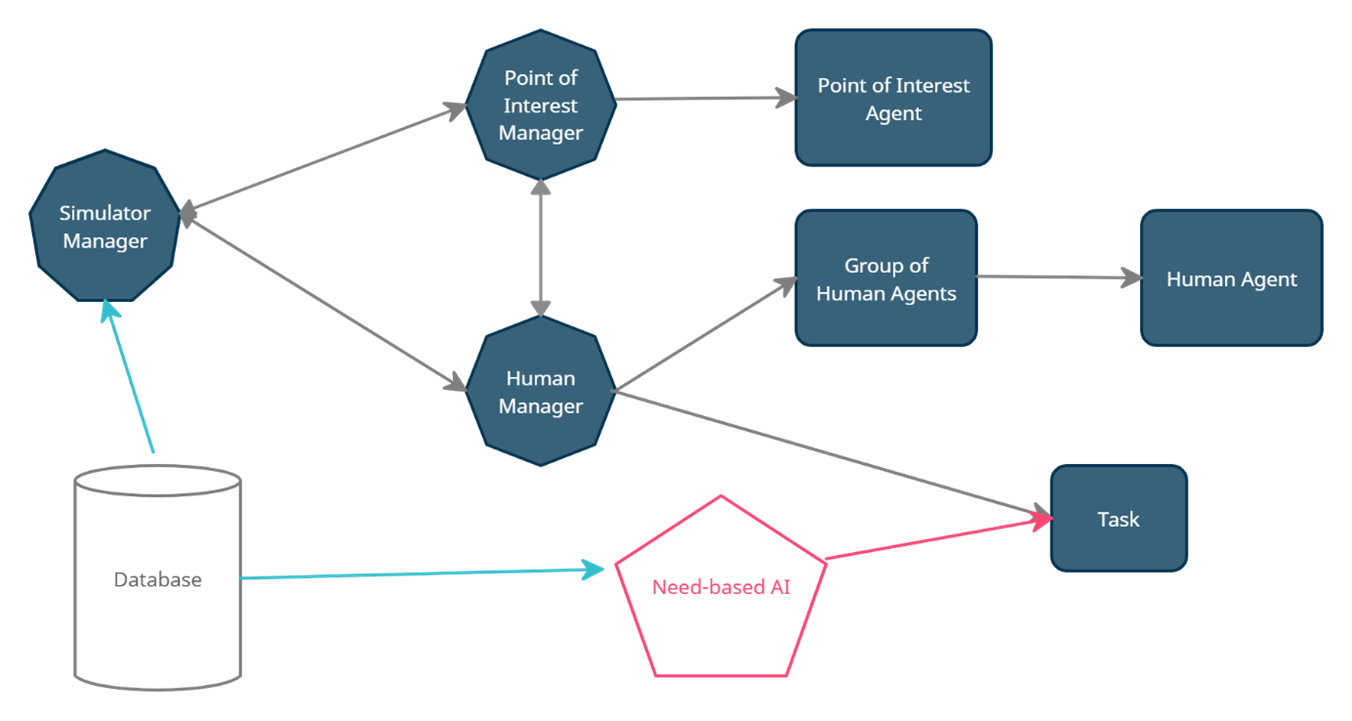}
\caption{Model architecture for COVID-CBABM}
\label{fig:mod-arch}
\end{figure}

In COVID19-CBABM, we have multiple agents interacting with various POIs, based on different parametric, preprogrammed behaviors. The parameter selection criteria is explained in Section 3.4. Figure \ref{fig:mod-arch} depicts the architecture of the system we developed. The model has four main components: Database, Simulator Manager, POI Manager, and Human Manager. The Database contains the parameter values. The Simulator Manager starts, controls, and ends the simulation. The POI Manager helps the Simulator Manager generate and update the parameters of the POI agents. The Human Manager helps the Simulator Manager generate and manage the Human agents.

\subsection{Simulation of Time}

The model is designed in steps, with each step or iteration being the minimum unit of time an agent can perform an action, such as moving to the adjacent grid square. Based on the total real-world time period being simulated, each iteration can represent from 1 hour to 2 hours of real-world time. A day in virtual time is equal to 12 iterations.

When the model starts the simulation, each Individual agent receives a daily schedule made of tasks. Each task has a start and end time between 0.0 and 24.0. Many tasks can happen within 2 hours or one iteration. Some tasks can require four iterations to finish. The main model has a variable named clock to keep track of daily hours during simulation so that an Individual agent can do tasks in the order of start time.

Each time the model finishes 12 steps, it updates the number of Individual agents, the associated parameters, and resets the clock variable to 0. 

\subsection{Simulation of Space}

The model uses available boundary information and generates neighborhood areas from the geojson shapefile. Individual agents can move both within a neighborhood area and between neighborhood areas.   Instead of placing the POIs randomly in the neighborhood areas, our model utilizes a geojson shapefile of Bronx borough with information about POIs within its boundary [5].

\begin{table}[]
\caption{Sample starting parameters for simulation run}
\label{tab:init-par}
\resizebox{\columnwidth}{!}{%
\begin{tabular}{|l|l|}
\hline
\textbf{Parameter}                           & \textbf{Value} \\ \hline
Number of Human agents                       & 10000          \\ \hline
Number of POI Agents                         & 4000           \\ \hline
Weather Factor W                             & 0.25           \\ \hline
Healthcare quality Factor P                  & 0.5            \\ \hline
Days of Simulation                           & 30             \\ \hline
Exposure distance                            & 100            \\ \hline
Mobility range                               & 10000          \\ \hline
Probability of Initial infected Population   & 0.01           \\ \hline
Probability of initial vaccinated Population & 0.575          \\ \hline
\end{tabular}%
}
\end{table}

\subsection{Agent Categories}

The current version of SEIR-CABM uses two different types of agents:

\subsubsection{Individual Agents}

Individual agents represent actual people. Their internal state is one of the SEIHRD compartments, with their modes of moving from one state to another being governed by the following parameters:

\begin{enumerate}
    \item \textit{SEIHRD State}: The current state of an individual agent: ‘S’, ‘E’, ‘I’, ‘H’, ‘R’, ‘D’. The movement between the states is governed by the POIs they have visited, and not by direct interactions between each other. 
    \item \textit{Profession}: The occupation of an individual can affect the rate of COVID-19 infection.
    \item \textit{Susceptibility}: The probability indicating how likely an agent is to become infected when exposed to an already infected agent. This is dependent upon age, profession, gender, and precautions taken within a specific POI. This value lies between 0.0 and 1.0, where 0.0 represents the ideal case in which everyone is locked down in quarantine, whereas the value 1.0 indicates no social distancing is considered. 
    \item \textit{Home location}: This is the quadrant of the grid where the agents originate from and return to after visiting their quota of POIs. This gives us some amount of control over the movement of the agents and by spacing them out, we prevent overcrowding of central POIs.

\end{enumerate}

We collect the total number of agents in each SEIHRD state at every iteration. For movement rules, the individuals select the closest POI agent to them. If there are multiple, they randomly select one. Then, they follow the shortest path to it, all while specific interaction rules guide them.

\subsubsection{Points of Interest Agents}

These are the locations visited by individual agents within the city of Houston, as determined by date from Safegraph \cite{noauthor_safegraph_nodate}. While the original dataset also has brand names, we anonymize those to remove company-specific biases and merge them into distinct categories. Each POI agent has the following parameters:

\begin{enumerate}
    \item \textit{Activity Period}: The period of time during which a POI is active. This represents the time of the day when individual agents are allowed to interact with the POI. For instance, if each model iteration is 4 hours, we have 6 iterations per day. We assign each POI a subset of time slots which range between 1 to 6 based on the window of the day they are covering. Other agents can interact with that POI only when the current internal simulation time matches the POI time.
    \item \textit{POI Occupancy}: This is the maximum number of agents that a POI is allowed to interact with every day. Once the quota is full, this POI is considered closed. All new agents coming to this POI will just pass through.
    \item \textit{POI Spread Probability}: This is the probability that, out of the total number of agents moving within 1 grid square of that particular POI, how many will be getting infected. So, as an example, a POI with low rates of commingling will have a value of 0.3 (30\% of agents infected) while those where people are faced into close quarters will have high rates of around 0.9 (90\% of all agents will be infected). CDC data is used to determine POI spread probabilities
\end{enumerate}

\subsection{Movement Rules}

While moving across the grid, all individual agents must follow certain rules to maintain the integrity of the simulation. These are:

\begin{enumerate}
    \item All Individual agents must go to work, school, or hospital daily at pre-decided POIs. The Individual agents will be assigned into certain groups due to their profession, and they will visit daily specific POI agents to conduct professional tasks.
    \item Each individual agent will follow schedules that were generated for them.
    \item When an individual agent visits POIs for service purposes, they must select the nearest POIs at every iteration. If there are multiple choices, they choose one at random.
    \item All Individual agents must attempt to move to their chosen POI by simply going straight to the POI if the distance is within the mobility range of Human agents. 
    \item If the distance to the chosen POI is outside of the mobility range of an Individual agent, he will use home POI as middle point. The Individual agent chooses the middle points with the shortest path.
    \item All Individual agents return home at the end of the day, except H agents and D agents.
    \item D agents and H agents do not move around.
    \item When an Individual agent is an H agent, he immediately goes to a hospital POI agent inside the city.
    \item Having been sick for over seven days, they become stationary.
    \item All Individual agents do not travel outside the city.
    \item All Human agents in a profession perform similar tasks daily in a pre-planned schedule.
    
\end{enumerate}
   
\subsection{Parameter Selection}

The initial number of agents is a representative sample of the city population. We set the number of POIs from the ratio of the total number of people visiting tracked locations to the total number of tracked locations during an average day (averaging all the days within a month), as determined by the Safegraph Weekly Places Patterns dataset.

\subsubsection{Initial Values for POI}
The value of $\beta$ (average contact rate) within the city of Houston and the mean infectious period are obtained from the CDC and the Texas Health and Human Services.
For susceptibility, we use a variety of sources. The age distribution is determined using Census data. Social Distancing is calculated from the Safegraph Social Distancing Metrics dataset, which has aggregated daily views of foot traffic between census block groups. 

\subsubsection{Initial Values for Individual Agents}
Activity period and POI quota are collected from Safegraph data. Disease spread probability comes from the CDC rates of COVID-19 infection. The activity period is determined by taking the time periods when the population within that POI was the largest and set that as their active times. While this does exclude non-peak times, we assume their effect of spreading infectious disease is minimal. Having a larger number of iterations in the model will give us more fine-grained control over the active time periods.
As the optimal parameter selection is NP-hard, we aim for a local optimum. Currently, we use a simple stochastic hill-climbing algorithm to enhance initial agent parameter selections. In this approach, we modify a single parameter at random and evaluate the current model parameters with a fitness function listed in (1). If the modification leads to an improvement in fitness then, we continue in that direction modifying another parameter otherwise, we undo the change of the parameter and we choose another parameter at random to be modified. This process is repeated over several iterations. 

For our fitness function, we minimize the Root Mean Square Error between the SEIR rates determined by our model and the actual SEIR states as reported by the Texas HHS. Let the total number of infections over our study period as determined by our model be $T_e$, and the actual official number of infections over the same period be $T_a$. Let our total study period be $N_d$ days. Then, our RMSE is:
\begin{equation}
RMSE = \sqrt{\frac{1}{N_{d}}*\sum_{i=1}^{N_d}\left( T_a - T_e \right)^{2}}
\end{equation}

The goal of our algorithm is to minimize this function over full 250 batches and for all 4 levels of severity. This gives us an ideal set of parameters covering a variety of conditions and improves the predictive powers of this model.

\subsection{Model Portability}

The key feature of our model is its ability to be reused over multiple cities and different time periods. This is a three step process. 

\subsubsection{Change Basemap} 
As a first step, we utilize a publicly available basemap layer e.g. from OpenStreetMap \cite{OpenStreetMap}, and update it to match the city to be ported. This basemap layer must contain city boundaries and road networks built in, to enable realistic agent movement.
\subsubsection{Modify Parameters}
Next, we update all model parameters to fit the new city. The infection rates, weather conditions, vaccination status, etc. are all obtained from publicly available CDC reports online and are automatically parsed into the model. The starting parameters may need to be calibrated with past data in order to achieve the most realistic results.
\subsubsection{Update Movement Data} 
For accurate Points-of-Interest data, we need the movement data from SafeGraph for the newer city. Safegraph is available online and can be pulled into the model through an API. However, this breaks if Safegraph modifies their database schema and must be verified for correctness every time.

\begin{figure}[t]
\centering
\includegraphics[scale=0.35]{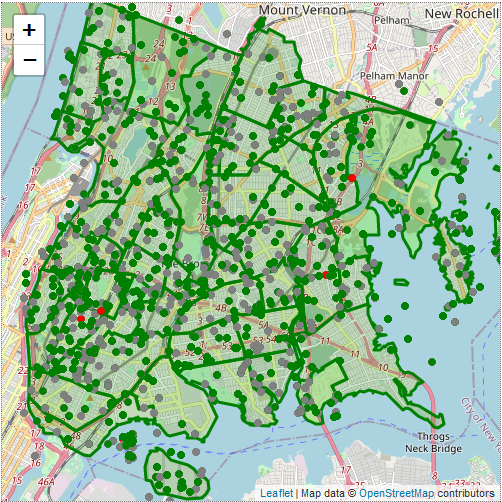}
\caption{COVID-CBABM Model State on Day 1}
\label{fig:day1}
\end{figure}

\section{Simulation Model}

\subsection{Dataset}

As the spread of the COVID-19 disease is highly dependent on people’s interactions with each other and their mobility, they must be integrated into the model of simulation. Regarding this fact, we use the Safegraph dataset [27]. SafeGraph is a data company that aggregates anonymized location data from numerous applications to provide insights about physical places. The dataset used provides unique and valuable insights into how people interact with their surroundings during this pandemic particularly foot-traffic to businesses and consumer points-of-interest. The core places module contains base information such as location name, address, category, and brand association for points of interest (POIs) where people spend time or money. It is available for approximately 6 million POI in the United States. The second module is a dataset of the visitor/place traffic and demographic aggregations that answer: how often people visit, how long they stay, where they came from, where else they go, and more. As people are engaging in social distancing, we use the Social Distancing Metrics module which gives some details about individuals staying at home and others traveling a specific distance from home. 

\subsection{Mesa-Geo Evaluation}

Our simulation was implemented using Python3 programming language. The main agent-based modeling framework we used was Mesa-geo \cite{noauthor_mesa-geo_2022}, after evaluating several potential solutions including GAMA and SPADE. While GAMA allowed better integration of Geospatial techniques, it is not adept at wrangling data. We preferred to use Python-based libraries due to the ease of usage in automating parameter extraction. We chose MESA over SPADE due to the ease of prototyping and very quick deployment, letting us able to iterate quickly. Before the simulation begins, Mesa-Geo uses an available geojson shapefile to create a virtual map of the city. The Mesa-Geo framework utilizes R-trees to efficiently compute queries and store spatial data indexes \cite{hunter_open-data-driven_2018}. It makes it possible for Human agents in the model to move in the simulated environment and interact. The built-in functions in Mesa-Geo allow Human agents to move quickly on the grid system by updating location information of the internal coordinate system, which conveniently compute the Euclidean distance between two agents on the simulated map.

\begin{figure}[t]
\centering
\includegraphics[scale=0.35]{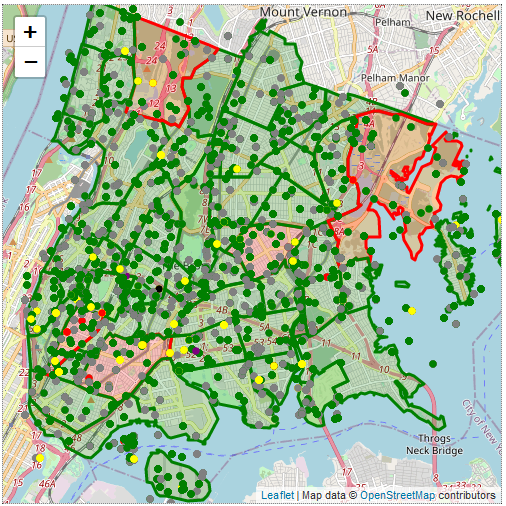}
\caption{COVID-CBABM Model State on Day 15}
\label{fig:day15}
\end{figure}

\begin{figure}[t]
\centering
\includegraphics[width=\columnwidth]{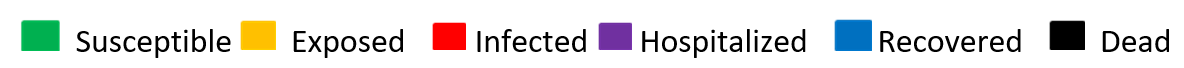}
\caption{Colour Labels for maps}
\label{fig:label}
\end{figure}

\subsection{Area of Interest}

We chose to run the model on the boroughs of New York because of the abundance of both demographic and geospatial data, and the ease of portability between them. We chose Brooklyn for the model. The borough-level data of COVID-19 spread in New York City is recorded and updated in many Github repositories. We selected Github repository nychealth as the main dataset, which contained summaries of total deaths, total hospitalized COVID-19 patients, total infected COVID-19 patients, total recovered people, and total vaccinated cases. In the directory nychealth/coronavirus-data, there are comma-separated values (CSV) files that record daily death cases, infected cases, hospitalized cases, and recovered cases. Those records are valuable actual data for us to extract parameters to calculate Protection Level and Resistance Level of Human Agents. For example, a CSV file could record death rate, infection rate, and hospitalized rate of COVID-19 victims in New York City by age. For each age group, we simply extracted the rates from the file and divided them by 100000 to get percentage values. We did the same for Gender and Income Level of COVID patients in New York City. 

\subsection{Model Access}

We present a working model for COVID-19 spread in the Bronx, New York City for December 2021. The model was deployed on a local machine running on Windows 10 with an Intel i9-9880H processor, 32 GB of DDR4 RAM and an NVMe drive with 8.0 GT/s, taking about 2 hours per run. Initial parameters for a run of the model can be seen in Table \ref{tab:init-par}. Deployment procedure for the model can be found in the github repository at \cite{rons_covid-cbabm_2022}. Simulation results for day 1 and day 15 can be seen in Figure 2 and Figure 3. The number of agents are kept low for a faster simulation time, but can be increased for greater accuracy.

% Please add the following required packages to your document preamble:
% \usepackage{graphicx}

\section{Conclusion}

In this paper, we have presented SEIR-CABM, a new ABM and SEIR inspired city-based modeling framework to analyze the dynamics of the COVID-19 in the borough of Brooklyn, New York City. Although it is challenging to simulate the actual virus, SEIR-CABM emulates the virus spread using an agent-based model to capture people’s interaction with their surroundings —in particular at different POIs. Our model uses the Safegraph data to realistically emulate the mobility of individuals and the chances of individuals to become infected when visiting particular types of POIs. To speed up this process, we developed automated Safegraph knowledge extraction procedures to obtain “realistic”, initial parameters for individual agents as well as POI agents.

Our model has the capability to simulate different scenarios of policies including the use of face masks, social distancing, and quarantine (stay at home) as well as their change based on events such as highly increasing and dropping infection rates. SEIR-CABM can help policymakers to evaluate different prevention policies by running SEIR-CABM simulating those scenarios and then comparing the obtained results. Although our model is tested using the NYC data, it is portable as it can be used for other cities:  the developed knowledge extraction procedures to Safegraph, CDC, and other datasets can be reused just using the data for that city.

As this is an ongoing work, we are planning to compare our model's prediction results with Covasim, extend our model by adding other types of agents, incorporating more details about people’s trajectories and schedules, letting agents have preferences in which POIs they regularly visit, improving the grid to more accurately resemble the city and exploring more in detail how the prevention policies add more insight on simulating the dynamics of the COVID-19 outbreak.

\bibliographystyle{ACM-Reference-Format}
\bibliography{sample-base}

%%% -*-BibTeX-*-
%%% Do NOT edit. File created by BibTeX with style
%%% ACM-Reference-Format-Journals [18-Jan-2012].

\begin{thebibliography}{7}

%%% ====================================================================
%%% NOTE TO THE USER: you can override these defaults by providing
%%% customized versions of any of these macros before the \bibliography
%%% command.  Each of them MUST provide its own final punctuation,
%%% except for \shownote{}, \showDOI{}, and \showURL{}.  The latter two
%%% do not use final punctuation, in order to avoid confusing it with
%%% the Web address.
%%%
%%% To suppress output of a particular field, define its macro to expand
%%% to an empty string, or better, \unskip, like this:
%%%
%%% \newcommand{\showDOI}[1]{\unskip}   % LaTeX syntax
%%%
%%% \def \showDOI #1{\unskip}           % plain TeX syntax
%%%
%%% ====================================================================

\ifx \showCODEN    \undefined \def \showCODEN     #1{\unskip}     \fi
\ifx \showDOI      \undefined \def \showDOI       #1{#1}\fi
\ifx \showISBNx    \undefined \def \showISBNx     #1{\unskip}     \fi
\ifx \showISBNxiii \undefined \def \showISBNxiii  #1{\unskip}     \fi
\ifx \showISSN     \undefined \def \showISSN      #1{\unskip}     \fi
\ifx \showLCCN     \undefined \def \showLCCN      #1{\unskip}     \fi
\ifx \shownote     \undefined \def \shownote      #1{#1}          \fi
\ifx \showarticletitle \undefined \def \showarticletitle #1{#1}   \fi
\ifx \showURL      \undefined \def \showURL       {\relax}        \fi
% The following commands are used for tagged output and should be
% invisible to TeX
\providecommand\bibfield[2]{#2}
\providecommand\bibinfo[2]{#2}
\providecommand\natexlab[1]{#1}
\providecommand\showeprint[2][]{arXiv:#2}

\bibitem[noa(2022a)]%
        {noauthor_mesa-geo_2022}
 \bibinfo{year}{2022}\natexlab{a}.
\newblock \bibinfo{title}{Mesa-{Geo}: a {GIS} extension for the {Mesa} agent-based modeling framework in {Python}}.
\newblock
\newblock
\urldef\tempurl%
\url{https://github.com/projectmesa/mesa-geo}
\showURL{%
\tempurl}
\newblock
\shownote{original-date: 2017-09-14T09:56:33Z}.


\bibitem[noa(2022b)]%
        {noauthor_safegraph_nodate}
 \bibinfo{year}{2022}\natexlab{b}.
\newblock \bibinfo{title}{SafeGraph Docs}.
\newblock
\newblock
\urldef\tempurl%
\url{https://docs.safegraph.com/docs}
\showURL{%
\tempurl}


\bibitem[Cuevas(2020)]%
        {cuevas_agent-based_2020}
\bibfield{author}{\bibinfo{person}{Erik Cuevas}.} \bibinfo{year}{2020}\natexlab{}.
\newblock \showarticletitle{An agent-based model to evaluate the {COVID}-19 transmission risks in facilities}.
\newblock \bibinfo{journal}{\emph{Computers in Biology and Medicine}}  \bibinfo{volume}{121} (\bibinfo{date}{June} \bibinfo{year}{2020}), \bibinfo{pages}{103827}.
\newblock
\showISSN{1879-0534}
\urldef\tempurl%
\url{https://doi.org/10.1016/j.compbiomed.2020.103827}
\showDOI{\tempurl}


\bibitem[Hunter et~al\mbox{.}(2018)]%
        {hunter_open-data-driven_2018}
\bibfield{author}{\bibinfo{person}{Elizabeth Hunter}, \bibinfo{person}{Brian Mac~Namee}, {and} \bibinfo{person}{John Kelleher}.} \bibinfo{year}{2018}\natexlab{}.
\newblock \showarticletitle{An open-data-driven agent-based model to simulate infectious disease outbreaks}.
\newblock \bibinfo{journal}{\emph{PLOS ONE}} \bibinfo{volume}{13}, \bibinfo{number}{12} (\bibinfo{date}{Dec.} \bibinfo{year}{2018}), \bibinfo{pages}{1--35}.
\newblock
\urldef\tempurl%
\url{https://doi.org/10.1371/journal.pone.0208775}
\showDOI{\tempurl}
\newblock
\shownote{Publisher: Public Library of Science}.


\bibitem[{OpenStreetMap contributors}(2022)]%
        {OpenStreetMap}
\bibfield{author}{\bibinfo{person}{{OpenStreetMap contributors}}.} \bibinfo{year}{2022}\natexlab{}.
\newblock \bibinfo{title}{{Planet dump retrieved from https://planet.osm.org }}.
\newblock \bibinfo{howpublished}{\url{ https://www.openstreetmap.org }}.
\newblock


\bibitem[RonS(2022)]%
        {rons_covid-cbabm_2022}
\bibfield{author}{\bibinfo{person}{RonS}.} \bibinfo{year}{2022}\natexlab{}.
\newblock \bibinfo{title}{{COVID}-{CBABM}: {A} {City}-{Based}, {Agent}-{Based} {Model} for simulating {Covid}-19 infections}.
\newblock
\newblock
\urldef\tempurl%
\url{https://github.com/RaunakDune/covid-cbabm}
\showURL{%
\tempurl}
\newblock
\shownote{original-date: 2022-09-01T13:42:58Z}.


\bibitem[Van Dyke~Parunak et~al\mbox{.}(1998)]%
        {van_dyke_parunak_agent-based_1998}
\bibfield{author}{\bibinfo{person}{H. Van Dyke~Parunak}, \bibinfo{person}{Robert Savit}, {and} \bibinfo{person}{Rick~L. Riolo}.} \bibinfo{year}{1998}\natexlab{}.
\newblock \showarticletitle{Agent-{Based} {Modeling} vs. {Equation}-{Based} {Modeling}: {A} {Case} {Study} and {Users}’ {Guide}}. In \bibinfo{booktitle}{\emph{Multi-{Agent} {Systems} and {Agent}-{Based} {Simulation}}} \emph{(\bibinfo{series}{Lecture {Notes} in {Computer} {Science}})}, \bibfield{editor}{\bibinfo{person}{Jaime~Simão Sichman}, \bibinfo{person}{Rosaria Conte}, {and} \bibinfo{person}{Nigel Gilbert}} (Eds.). \bibinfo{publisher}{Springer}, \bibinfo{address}{Berlin, Heidelberg}, \bibinfo{pages}{10--25}.
\newblock
\showISBNx{978-3-540-49246-7}
\urldef\tempurl%
\url{https://doi.org/10.1007/10692956_2}
\showDOI{\tempurl}


\end{thebibliography}

\end{document}